\documentclass[12pt]{article}
\usepackage{graphicx}
\usepackage{graphicx}
\usepackage{amsfonts}
\usepackage{amssymb}
\usepackage{amsmath}

\newlength{\dinwidth}
\newlength{\dinmargin}
\def\lapproxeq{\lower .7ex\hbox{$\;\stackrel{\textstyle
<}{\sim}\;$}}
\def\gapproxeq{\lower .7ex\hbox{$\;\stackrel{\textstyle
>}{\sim}\;$}}

\def\bea{\begin{eqnarray}}
\def\eea{\end{eqnarray}}

\def\beq{\begin{equation}}
\def\eeq{\end{equation}}

\def\qbold{\mbox{\boldmath${q}$}}

\begin{document}
\titlepage
\vspace*{1in}
\begin{center}
{\Large \bf  Resummation effects in  Higgs boson transverse momentum distribution
within the framework of unintegrated parton distributions}

\vspace*{0.4in}
Agnieszka Gawron and \fbox{Jan Kwieci\'nski}
\\
\vspace*{0.5cm}

{\it The H. Niewodnicza\'nski Institute of Nuclear Physics \\
 Radzikowskiego 152, 31-342 Cracow, Poland} \\
\end{center}
\vspace*{1cm}

\vskip1cm
\begin{abstract}
The cross sections describing the transverse momentum distributions of Higgs bosons
 are discussed within the framework of unintegrated parton distributions
obtained from the  CCFM equations in
the single loop approximation.  It is shown how the approximate treatment of the CCFM equations
generates the standard expressions describing the soft gluon resummation effects in the corresponding
cross sections.  Possible differences between exact and approximate solutions of the
CCFM equations are discussed on the example of the $gg\rightarrow H$ fusion mechanism,
which gives the dominant contribution to  Higgs production.
\end{abstract}

\newpage

\section{Introduction}

The transverse momentum distributions of Drell-Yan pairs, the $W$ or $Z_0$ gauge bosons
or the Higgs bosons  produced in high energy hadronic collisions  are known to
be strongly affected  by the soft gluon
(recoil) resummation effects \cite{ESW} - \cite{WJSKUL}.  These effects are particular
important in the region
$p_T^2\ll Q^2$, where $Q$ and $p_T$ denote   the mass and transverse momentum
of the virtual photon $\gamma^*$ or  the $W, Z_0$ or Higgs bosons, respectively.
The soft gluon resummation effects are also important in the prompt photon production in hadronic
collisions.

All these reactions   reflect the basic partonic fusion subprocesses, i.e. $q + \bar q \rightarrow
\gamma^* \rightarrow l^+l^-$, $q + \bar q \rightarrow
W$, $q + \bar q \rightarrow Z_0$, $g+g\rightarrow H$, $g+ q (\bar q) \rightarrow q (\bar q)
+\gamma$ etc.  The  transverse momenta of the lepton pairs, the $W, Z_0$ and/or
Higgs bosons etc. should therefore just be equal to the sum of the transverse momenta of the
colliding partons.   The transverse momentum distributions of the lepton pairs, the $W, Z_0$ and/or
Higgs bosons etc. produced through the fusion of partons do  therefore probe the
transverse momentum distributions of the partons, i.e.  the {\it unintegrated} parton
distributions \cite{STERMAN,GIULIA},\cite{KMR1}-\cite{COLLINS}.
The corresponding cross sections are determined by the
$k_T$ factorisation - like  convolutions (subject
to the transverse momentum conservation) of the unintegrated parton distributions with the  corresponding
partonic cross sections.   In this simple extension of the collinear picture to the more
exclusive
configuration,  in which the transverse momenta are not integrated over the
gluon resummation effects,
are  naturally attributed to the unintegrated parton distributions.
     Evolution of the unintegrated
distributions is described in perturbative QCD by the Catani, Ciafaloni, Fiorani, Marchesini (CCFM)
equation \cite{CCFM}.  It corresponds to the sum of the ladder diagrams with the angular
ordering along the chain. The   gluon radiation controlling evolution of
the unintegrated parton distributions
 generates the transverse momentum
of the  partons through the simple recoil effect.  The CCFM equation interpolates between
the (LO) DGLAP evolution at large and moderately small values of $x$  and the BFKL
dynamics at low $x$.   Pure DGLAP evolution corresponds to the so called ``single loop approximation''
of the CCFM evolution \cite{CCFM}-\cite{GM2}.

The purpose of this paper is to examine the transverse momentum
distributions of the Higgs bosons produced in hadronic collisions
using the unintegrated parton distributions framework. To be precise  we shall limit ourselves to
the $gg\rightarrow H$
fusion mechanism and determine the corresponding cross sections in terms of the
unintegrated gluon distributions.    The latter
will be  obtained from the solution of the CCFM equation in the single loop approximation, which
may be adequate for the moderately small values of $x$ probed in the (central) Higgs boson
 production.
We shall utilise the fact that this equation can be diagonalised by the Fourier--Bessel
transform and becomes evolution
equation for the (scale dependent) parton distributions $\bar f_i(x,b,Q)$ for fixed $b$, where $b$ is a vector
of impact parameter
conjugate to the transverse momentum.  We express the cross sections in terms of the
function $f_G(x_{1,2},b,Q)$ and show that the CCFM equation correctly reproduces
the double and single  $\ln(Q^2b^2)$ resummation in the limit $Q^2b^2\gg 1$, that corresponds
to $p_T^2\ll Q^2$.

The content of our paper is as follows.  In the next section we recall
the basic formulas describing the soft gluon resummation effects in the transverse
momentum distributions describing of Higgs bosons
produced through the gluon-gluon fusion mechanism.  In section 3 we discuss this distribution
within the unintegrated gluon distributions framework.  We point out that the basic quantities
controlling the transverse momentum distributions in this framework are the unintegrated
distributions in  the $b-$representation.  In section 4 we discuss the CCFM equation
in this representation.  Section 5 contains discussion of the soft gluon resummation
formulas as the result of the approximate treatment of the solution(s) of the CCFM equation in
the $b-$representation.   In section 6 we present numerical results concerning the
transverse momentum distributions of Higgs bosons and compare predictions based
upon exact solution of the CCFM equation with the standard LO soft gluon resummation
formulas.  Finally, in section 7 we give summary of our results.

\section{Soft gluon resummation formulas}

The soft gluon resummation effects are most conveniently formulated  using the 
$b$-representation.  Thus the cross-section for Higgs boson production in $p \bar p$
collisions corresponding to the gluon - gluon fusion subprocess
$gg \rightarrow H$ is expressed in the following way \cite{BERGER,WJSKUL}
\beq
 {\partial \sigma\over \partial Q^2 \partial y \partial p_T^2}\,=\,
{\sigma_0\, Q^2\over \tau s}\,  \pi^2 \delta (Q^2-m_H^2)\,
{1\over 2}\int_0^{\infty}db\, b\, J_0(bp_{T})\, W_{GG}(x_1,x_2,b,Q),
\label{sighb1}
\eeq
with $J_0(z)$ being the Bessel function.
In  eq.~(\ref{sighb1}), ${p_T}$ and $y$ denote the transverse momentum and rapidity of the
Higgs boson, respectively, and $m_H$ is the Higgs boson  mass while $x_{1,2}$ are the
 longitudinal momentum  fractions of the gluons given by
\bea
\nonumber
x_1\!\!&=&\!\!\sqrt{{Q^2+p_T^2\over s}}\,\exp(y)
\\
 x_2\!\!&=&\!\!\sqrt{{Q^2+p_T^2\over s}}\,\exp(-y)
\label{xab}
\eea
where $s$ is the square of the CM energy of the colliding $p \bar p$ system,
and
\beq
\tau \,=\, x_1 x_2\,.
\label{tau}
\eeq
The cross section $\sigma_0$ is given by:
\beq
\sigma_0={\sqrt{2}\, G_F\over 576\, \pi}\,\alpha_S(\mu_r^2)
\label{sigma0}
\eeq
where $G_F$ is the Fermi constant.   The corresponding cross-section for the Drell-Yan
production of muon pairs reads:
\beq
{\partial \sigma \over \partial Q^2 \partial y \partial p_T^2}
\,=\, {\sigma^{DY}_0\over \tau}\,{1\over 2}\int db\, b\,
J_0(bp_T)
\sum_i e_i^2\left\{W_{\bar q_i q_i}(x_1,x_2,b,Q)+ W_{q_i \bar q_i}(x_1,x_2,b,Q)\right\}
\label{csxdyb1}
\eeq
where now  ${p_T}, Q$ and $y$ denote the transverse momentum, invariant mass and
rapidity of the lepton pair, respectively, and $x_{1,2}$ denote the momentum fractions of
the parent
hadrons carried by a quark or antiquark and are defined by eq.~(\ref{xab}).
The elementary DY cross section $\sigma^{DY}_0$ is given by:
\begin{equation}
\sigma^{DY}_0={4\pi \alpha_{em}^2\over 9 s Q^2}\,.
\label{sigmady0}
\end{equation}
The basic functions driving the corresponding cross sections
are the functions $W_{ij}(x_1, x_2,b,Q)$ which
can be interpreted as  the scale dependent parton luminosities in the $b-$representation.
The soft gluon resummation gives the following representation of these functions
\bea
W_{ij}(x_1,x_2,b,Q)\!\!&=&\!\!
W_{ij}^{NP}(x_1, x_2,b,Q)\,\exp\left\{{1\over 2}\left[S_{i}(b,Q)+S_j(b,Q)\right]
\right\}
\nonumber \\
&\times&\!\!
x_1\, p_i^{eff}(x_1,c^2/b^2)\,x_2\,p_j^{eff}(x_2,c^2/b^2).
\label{swij}
\eea
The functions $p_{i}^{eff}$
are given by the conventional integrated parton distributions $p_i$ probed at the scale
$\mu^2=c^2/b^2$ with $c \sim 1$ \cite{CSS}, i.e.
\beq
 p_{i}^{eff}(x,\mu^2)\, =\, p_{i}(x,\mu^2)\,+\,O(\alpha_s(\mu^2))
\label{pieff}
\eeq
 with the $O(\alpha_s(\mu^2))$ corrections given in \cite{DAVSTIR,ERV}.  The exponents $S_i$ are given by:
\beq
S_i(b,Q)\,=\,-\int_{c^2/b^2}^{Q^2}{d\bar \mu^2 \over \bar \mu^2}\left[\ln\left({Q^2\over \bar \mu^2}\right)
A_i(\alpha_s(\bar \mu^2))+  B_i(\alpha_s(\bar \mu^2))\right],
\label{sibq}
\eeq
where the functions  $A_i$ and $B_i$ are defined by the perturbative
expansion:
\bea
A_i(\alpha_s)\!\!&=&\!\!\sum_{n=1}^{\infty}\left({\alpha_s\over 2\pi}\right)^nA_i^{(n)}
\nonumber \\
B_i(\alpha_s)\!\!&=&\!\!\sum_{n=1}^{\infty}\left({\alpha_s\over 2\pi}\right)^nB_i^{(n)}\,.
\label{abn}
\eea
The LO coefficients are given by \cite{KODAIRA}:
\bea
A_G^{(1)}\!\!&=&\!\!2C_A
\nonumber \\
B_{G}^{(1)}\!\!&=&\!\!{4\over 3} N_f T_R-{11\over 3} C_A
\label{log}
\\
A_{q}^{(1)}\!\!&=&\!\!A_{\bar q}^{(1)}=C_F
\nonumber \\
B_{q}^{(1)}\!\!&=&\!\!B_{\bar q}^{(1)}=-{3\over 2}C_F
\label{loq}
\eea
where for the $SU(3)_c$ we have: $C_A=3, C_F=4/3, T_R=1/2$, and $N_f$ is the number of
active flavours.  Higher order coefficients are defined in \cite{FLORIAN}.
The factors $W_{ij}^{NP}(x_1,x_2,b,Q)$ describe the non-perturbative contribution(s) \cite{QIU,BERGER,WJSKUL}.

Coefficients $A_i^{(n)}$ and $B_i^{(n)}$ are directly connected to the part of the splitting
functions $P_{ij}(z)$ corresponding to real emission.
To this aim we introduce the functions $R_G(z)$ and $R_q(z)$
defined by:
\bea
R_G(z)\!\!&=&\!\!z[ P_{gg}(z)+ P_{qg}(z)]
\label{rg1}
\\
R_q(z)\!\!&=&\!\!z[ P_{qq}(z)+ P_{gq}(z)]\,,
\label{rq1}
\eea
and rearrange these functions as below
\bea
 R_{G}(z)\!\!&=&\!\!{r_G\over 1-z}\, +\, \bar R_G(z)
\label{rg2}
\\
R_{q}(z)\!\!&=&\!\!{r_q\over 1-z}\, +\, \bar R_q(z)
\label{rq2}
\eea
where $r_q=C_F$ and $r_G=2C_A$, and
$R_G(z)$ and $R_q(z)$ are regular at $z=1$.  In LO we get the following relations
\bea
A_i^{(1)}\!\!&=&\!\!r_i\,,~~~~~~~~~~~~~~~~~i=q,G
\label{airi1}
\\
B_i^{(1)}\!\!&=&\!\!2\int_0^1 dz\, \bar R_i(z)\,.
\label{biri1}
\eea
It is also useful to remind the following  identity:
\beq
\int_0^1 dz\, [R_q(z)-P_{qq}(z)]\,=\,0\,.
\label{rqpqq}
\eeq

\section{Transverse momentum distributions of Higgs bosons and DY pairs}

The cross section for Higgs production in $p \bar p$ collisions corresponding to
the gluon--gluon fusion subprocess
$gg \rightarrow H$ can be expressed in the following way in terms of the scale dependent
unintegrated gluon distributions $f_G(x_{1,2},k_T,Q)$
\bea
 {\partial \sigma\over \partial Q^2 \partial y \partial p_T^2}\!\!&=&\!\!
{\sigma_0 Q^2\over \tau s}\,\pi \delta (Q^2-m_H^2)
\nonumber \\
&\times&\!\!
\int \frac {d^{2}{\bf k_{1}}\,d^2{\bf k_{2}}}{\pi^2}\,
  f_{G}(x_{1}, k_{1}, Q)\, f_{G}(x_{2}, k_{2}, Q) \,
\delta ^{(2)}({\bf k_{1}}+ {\bf k_{2}} - {\bf p_{T}})\,.
\label{sigmah1}
\eea
 In this equation ${\bf p_T}$ and $y$ denote, as before, the transverse momentum and rapidity of the
Higgs boson, respectively, $m_H$ is the Higgs boson  mass, ${\bf k_{1,2}}$ are the
transverse momenta of the colliding gluons and $x_{1,2}$ are their longitudinal momentum
fractions (see eq.~(\ref{xab})).
The cross-section $\sigma_0$ is given by eq.~(\ref{sigma0}).
From eqs.~(\ref{xab},\ref{tau}) we also have
\begin{equation}
\tau s \,=\, Q^2+p_T^2
\label{taus}
\end{equation}
Substituting in eq.~(\ref{sigmah1})  the following representation of the
$\delta$ function,
\beq
\delta ^{(2)}({\bf k_{1}}+ {\bf k_{2}} - {\bf p_{T}})\,=\,{1\over (2 \pi)^2}
\int d^2{\bf b}\, \exp\{-i({\bf k_{1}}+ {\bf k_{2}} - {\bf p_{T}}){\bf b}\}\,,
\label{delta2}
\eeq
we find that the integral defining the differential cross section for Higgs production
can be directly expressed in terms of the
unintegrated gluon distributions $\bar f_G(x,b,Q)$ in the impact parameter representation
\beq
 {\partial \sigma\over \partial Q^2 \partial y \partial p_T^2}\,=\,
{\sigma_0 Q^2\over \tau s}\, \pi^2\delta (Q^2-m_H^2)
\int_0^{\infty}db^2 \,\bar f_{G}(x_{1}, b, Q)\, \bar f_{G}(x_{2}, b, Q)\, J_0(bp_{T})\,.
\label{sigh2}
\eeq
where
\begin{equation}
\bar f_{G}(x,b,Q)\,=\,\int_0^{\infty}dk_T k_T\, J_0(bk_T)\,f_G(x,k_T,b)\,.
\label{barfg}
\end{equation}

The cross section for Higgs production which is driven by the $gg\rightarrow H$ fusion is determined
by the   gluon
distributions. Similar representation
of the cross section, but in terms of the quark and antiquark
distributions in the $b$ space, can be obtained
in the case of the Drell-Yan production of the muon pairs. The latter reaction reflects
the process $q \bar q \rightarrow \gamma^* \rightarrow \mu^+\mu^-$, and the corresponding
expression for the cross-section
reads
\beq
{\partial \sigma \over \partial Q^2 \partial y \partial p_T^2}\, =\, {\sigma^{DY}_0\over \tau} \int db^2 J_0(bp_T)
\sum_i e_i^2 \left[\bar f_{q_i}^{1}(x_1,b,Q)\,\bar f_{\bar q_i}^{2}(x_2,b,Q)\,+\,
\bar f_{q_i}^{2}(x_2,b,Q)\,\bar f_{\bar q_i}^{1}(x_1,b,Q)\right]
\label{csxdyb}
\eeq
where now  $p_T, Q$ and $y$ denote the transverse momentum, invariant mass and
rapidity of the lepton pair, respectively, $x_{1,2}$ denote the momentum fractions (\ref{xab}) of
the parent
hadrons carried by a quark or antiquark.
The functions $\bar f_{q_i}^{1,2}$  and $\bar f_{\bar q_i}^{1,2}$
are the quark and antiquark unintegrated distributions in the $b-$representation.
 As in the case for the
gluons, they are linked to
the scale dependent $k_T$ distributions of quarks and antiquarks through the Fourier--Bessel transform
 (cf. eq.~(\ref{barfg})).  The elementary DY cross section $\sigma^{DY}_0$ is given by eq.~(\ref{sigmady0}).
Comparing eqs.~(\ref{sigh2}, \ref{csxdyb}) with eqs.~(\ref{sighb1}, \ref{csxdyb1}),
we find that within  the unintegrated parton distrtibution framework the functions
$W_{ij}(x_1,x_2,b,Q)$ are given by:
\beq
W_{ij}(x_1,x_2,b,Q)\,=\,4\,\bar f_i(x_1,b,Q)\,\bar f_j(x_2,b,Q)\,.
\label{wij}
\eeq

\section{Single loop CCFM equation in the $b-$representation}

 In this section we recall the CCFM equation in the single loop approximation.
 Then we extend this equation to the system of equations incorporating
 unintegrated quark and antiquark distributions,
 $f_{q_i}$ and $f_{\bar q_i}$.
The  original Catani-Ciafaloni-Fiorani-Marchesini
(CCFM) equation \cite{CCFM,GM1} for the unintegrated, scale-dependent gluon distribution
$f_G(x,k_T,Q)$ has the following form:
\bea
f_G(x,k_T,Q) \!\!&=&\!\! f_G^0(x,k_T,Q) \,+\, \int{d^2\qbold \over \pi q^2}
\int_x^1 {dz\over z}\, \Theta(Q-qz)\, \Theta(q-q_0)\, { \alpha_s\over 2\pi }\, \Delta_S (Q,q,z)
\nonumber \\
\nonumber \\
&\times&\!\!
\left[2N_c\,\Delta_{NS}(k_T,q,z)\, +\, {2N_c\, z\over 1-z}\right]
f\!\left({x\over z},|{\bf k_T}+(1-z)\qbold|,q \right),
\label{ccfm1}
\eea
where $\Delta _S(Q,q,z)$ and $\Delta_{NS}(k_T,q,z)$ are the Sudakov and non-Sudakov
form factors,
\bea
\Delta _S(Q,q,z)\!\!&=&\!\! \exp\left\{-\int_{(qz)^2}^{Q^2}{dp^2\over
p^2}{\alpha_s\over 2 \pi} \int _0^{1-q_0/p}dz\,z\,P_{gg}(z)\right\}
\label{ds}
\\ \nonumber
\\
\Delta_{NS}(k_T,q,z)\!\!&=&\!\! \exp\left\{-\int_z^1{dz'\over z'}
\int_{(qz')^2}^{k_T^2} {dp^2\over p^2}{2N_c\,\alpha_s\over 2 \pi}\right\} .
\label{dns}
\eea
The variables $x, k_T$ and $Q$ denote the longitudinal momentum fraction,
the transverse momentum of the gluon and the hard scale, respectively.
Eq.~(\ref{ccfm1}) is generated by the sum of ladder diagrams with  angular
ordering along the chain. The hard scale $Q$
is defined in terms of the maximum emission angle \cite{LUNDSMX,CCFM}.
The constraint $\Theta(Q-qz)$ in eq.~(\ref{ccfm1}) reflects the angular ordering,
and the inhomogeneous term,
$f^0(x,k_T,Q)$, is related to the input non-perturbative gluon distribution.
It also contains effects of both the Sudakov and non-Sudakov form-factors \cite{KMSU}.

Eq.~(\ref{ccfm1})  interpolates between
a part  of
the DGLAP evolution at large and moderately small values of $x$,
and the BFKL dynamics at small $x$.  To be precise, it contains only the $g\rightarrow gg$
splittings and only those  parts of the
splitting function $P_{gg}(z)$ which are singular at either $z \rightarrow 1$
or $z \rightarrow 0$. Following  \cite{AGJK,AGJKWB} we shall make the following
extension of the original CCFM equation (\ref{ccfm1}):
\begin{enumerate}
\item We introduce, besides the unintegrated gluon distributions $f_G(x,k_T,Q)$,
also the unintegrated quark and antiquark gluon distributions, $f_{q_i}(x,k_T,Q)$ and
 $f_{\bar q_i}(x,k_T,Q)$.
\item We include, in addition to the $g\rightarrow gg$ splittings, also  the $q\rightarrow gq$,
$\bar q \rightarrow g\bar q$, and $g \rightarrow
\bar q q$ transitions along the chain.
\item We take into account the complete splitting functions $P_{ab}(z)$, and not only their singular
parts.
\end{enumerate}
These extensions  make the CCFM framework more
accurate in the region of large
and moderately small values of $x$ (i.e. $x\ge 0.01$ or so).
In this region one can also    introduce the single-loop approximation
that corresponds to the following
replacements:
\bea
\Theta(Q-qz) &\longrightarrow& \Theta(Q-q)
\label{sloop1}
\nonumber \\
\Delta_{NS}(k_T,q,z) &\longrightarrow& 1. \label{sloop2}
\eea

It is also convenient  to ``unfold'' the Sudakov form factor such that
the real emission and virtual terms appear on an equal footing in the corresponding
evolution equations.  Finally, the unfolded system of the extended CCFM equations in the single-loop
approximation has the following form:
\bea
\nonumber
f_{NS}^i(x,k_T,Q)\!\!&=&\!\! f^{i0}_{NS}(x,k_T)\, +\int_0^1dz\int{d^2q\over \pi q^2}\, {\alpha_s(q^2)\over
2\pi}\,\Theta(q^2-q_0^2)\,\Theta(Q-q)
\\ \nonumber
\\
&\times&\!\! P_{qq}(z)\left[\Theta(z-x)\,f_{NS}^i\!\left({x\over z},k^{\prime}_T,q\right)-f_{NS}^i(x,k_T,q) \right]
\label{ccfmns}
\eea
\bea
\nonumber
f_{S}(x,k_T,Q)\!\!&=&\!\!f_S^0(x,k_T)\,+ \int_0^1dz\int{d^2q\over \pi
q^2}\, {\alpha_s(q^2)\over 2\pi}\,\Theta(q^2-q_0^2)\,\Theta(Q-q)
\\ \nonumber
\\ \nonumber
&\times&\!\! \bigg\{\Theta(z-x)\left[P_{qq}(z)\,f_{S}\!\left({x\over
z},k^{\prime}_T,q\right)+ P_{qg}(z)\,f_{G}\!\left({x\over
z},k^{\prime}_T,q\right)\right]
\\ \nonumber
\\
&-&\!\! P_{qq}(z)\,f_{S}(x,k_T,q)\bigg\}
\\ \nonumber
\\ \nonumber
f_{G}(x,k_T,Q)\!\!&=&\!\!f_G^0(x,k_T)\,+ \int_0^1dz\int{d^2q\over \pi q^2} {\alpha_s(q^2)\over
2\pi}\,\Theta(q^2-q_0^2)\,\Theta(Q-q)
\\ \nonumber
\\ \nonumber
&\times&\!\! \bigg\{\Theta(z-x)\left[P_{gq}(z)\,f_{S}\!\left({x\over
z},k^{\prime}_T,q\right)+ P_{gg}(z)\,f_{G}\!\left({x\over
z},k^{\prime}_T,q\right)\right]\bigg.
\\ \nonumber
\\
&-&\!\!  \bigg.
\big[zP_{gg}(z)+zP_{qg}(z)\big]\,f_{G}(x,k_T,q)\bigg\}
\label{ccfmg}
\eea
where
\begin{equation}
{\bf k^{\prime}_T} ={\bf k_T} + (1-z){\bf q} .
\label{qprimet}
\end{equation}
The functions $f_{NS}^i(x,k_T,Q)$ are the unintegrated non-singlet quark  distributions, while the
unintegrated singlet distribution, $f_{S}(x,k_T,Q)$, is defined as
\begin{equation}
f_S(x,k_T,Q)\,=\,\sum_{i=1}^{N_f}\, \left[f_{q_i}(x,k_T,Q)+ f_{\bar q_i}(x,k_T,Q)\right]\,.
\label{singlet}
\end{equation}
The functions $P_{ab}(z)$ are the LO splitting functions corresponding to real
emissions, {\em i.e.}:
\begin{eqnarray}
P_{qq}(z)\!\!&=&\!\!C_F\, {1+z^2\over 1-z} \nonumber \\
P_{qg}(z)\!\!&=&\!\!N_f\, [z^2+(1-z)^2] \nonumber \\
P_{gq}(z)\!\!&=&\!\!C_F\, {1+(1-z)^2\over z} \nonumber \\
P_{gg}(z)\!\!&=&\!\!2N_c\left[{z\over 1-z}+{1-z\over z} + z(1-z)\right]
\label{splitf}
\end{eqnarray}
where $N_f$ and $N_c$ denote the number of flavours and colours, respectively, and $C_F=(N_c^2-1)/(2N_c)$.
It should be observed that after integrating both sides of
eqs.~(\ref{ccfmns})-(\ref{ccfmg}) over $d^2{\bf k_T}$, we get the usual DGLAP equations for the
integrated parton distributions $p_k(x,Q^2)$, defined by
\begin{equation}
x\/p_k(x,Q^2)\,=\,\int_0^\infty dk_T^2\, f_k(x,k_T,Q).
\label{idis}
\end{equation}

A very important merit of eqs.~(\ref{ccfmns})-(\ref{ccfmg})
is the fact that they can be diagonalised by the Fourier-Bessel transform
\beq
f_{k}(x,k_T,Q)\,=\,\int_0^{\infty}db\, b\, J_0(b k_T)\,\bar f_{k}(x,b,Q) \label{fb1}\,,
\eeq
where the inverse relation reads
\beq
\bar f_{k}(x,b,Q)\,=\,\int_0^{\infty}dk_T \,k_T\, J_0(b k_T)\, f_{k}(x,k_T,Q)\,.
\label{fb2}
\eeq
In the above $k=NS,S,G$,  and $J_0$ is the Bessel function. At $b=0$ the
functions $\bar f_{k}(x,b,Q)$  are related to the integrated distributions $p_i(x,Q^2)$
\begin{equation}
\bar f_{k}(x,b=0,Q)\,=\,{1\over 2}\,x\/p_k(x,Q^2).
\label{fbxp}
\end{equation}

The corresponding
evolution  equations for $\bar f_{NS}$, $\bar f_{S}$ and $\bar f_{G}$,
which follow from eqs.~(\ref{ccfmns})-(\ref{ccfmg}), have the following form:
\bea
\nonumber
Q^2\,{\partial \bar f_{NS}(x,b,Q)\over \partial Q^2} \!\!&=&\!\!
{\alpha_s(Q^2)\over 2\pi}  \int_0^1dz  \,P_{qq}(z)
\bigg[\Theta(z-x)\,J_0((1-z)Qb)\,\bar f_{NS}\!\left({x\over z},b,Q\right)
\\ \nonumber
\\
&-&\!\!zP_{qq}(z)\,\bar f_{NS}(x,b,Q)
\bigg]
\label{dccfmnsb}
\eea
\bea\nonumber
Q^2\,{\partial \bar f_{S}(x,b,Q)\over \partial Q^2}
\!\!&=&\!\!
{\alpha_s(Q^2)\over 2\pi} \int_0^1 dz
\bigg\{\Theta(z-x)\,J_0((1-z)Qb)\bigg[P_{qq}(z)\,\bar f_{S}\!\left({x\over z},b,Q\right)
\\ \nonumber
\\
&+&\!\!
P_{qg}(z)\,\bar f_{G}\!\left({x\over z},b,Q\right)\bigg]
\,-\,[zP_{qq}(z)+zP_{gq}(z)]\,\bar f_{S}(x,b,Q)\bigg\}
\label{dccfmsb}
\eea
\bea\nonumber
Q^2\, { \partial \bar f_{G}(x,b,Q)\over \partial Q^2}
\!\!&=&\!\!
{\alpha_s(Q^2)\over 2\pi} \int_0^1 dz
\bigg\{\Theta(z-x)\,J_0((1-z)Qb)\bigg[P_{gq}(z)\,\bar f_{S}\!\left({x\over z},b,Q\right)
\\ \nonumber
\\
&+&\!\!
P_{gg}(z)\,\bar f_{G}\!\left({x\over z},b,Q\right)\bigg]
\,-\,[zP_{gg}(z)+zP_{qg}(z)]\,\bar f_{G}(x,b,Q)\bigg\}
\label{dccfmgb}
\end{eqnarray}
with the initial conditions
\begin{equation}
\bar f_k(x,b,Q_0)\,=\,\bar f_k^0(x,b)\,.
\label{bcond}
\end{equation}
We have therefore found that the CCFM scheme  in the single loop approximation does  directly
provide the system of evolution equations for the unintegrated
parton distributions $\bar f_k(x,b,Q)$ in the $b-$representations.  These
functions are the basic quantities  in  the
description of $p_T$ spectra, eqs.~(\ref{csxdyb1}),(\ref{sigh2}),
within the  unintegrated parton distributions framework.
In the next section  we discuss their solution and show that they correctly incorporate
the resummation effects.

\section{Soft gluon resummation formulas from the solution of the CCFM equations}

In order to understand the structure of  the solution of the
CCFM  equations (\ref{dccfmnsb})-(\ref{dccfmgb}),
it is convenient to introduce the moment functions $\tilde f_i(n,b,Q)$
\begin{equation}
\tilde f_i(n,b,Q)\,=\,\int_0^1dx\, x^{n-1}\bar f_i(x,b,Q)
\label{momf}
\end{equation}
We shall consider at first the non-singlet,  valence quark distributions for
which the CCFM equation in the moment space reads:
\bea
Q^2\,{\partial \tilde f_{NS}(n,b,Q)\over \partial Q^2}\, =\,
{\alpha_s(Q^2)\over 2 \pi}\int_0^1dz\, P_{qq}(z)\left\{z^nJ_0((1-z)Q\/b)-1\right\}\,\tilde f_{NS}(n,b,Q)
\label{emomns}
\eea
Its solution reads:
\begin{equation}
\tilde f_{NS}(n,b,Q)\,=\,\tilde f_{NS}(n,b,Q_0)\,\exp\{I(n,b,Q)\}
\label{smomns}
\end{equation}
where the function $I(n,b,Q)$ is given by:
\begin{equation}
I(n,b,Q)\,=\,\int_{Q_0^2}^{Q^2}{dq^{2}\over q^{2}}
{\alpha_s(q^{2})\over 2 \pi}\int_0^1 dz\, P_{qq}(z)\left\{z^nJ_0((1-z)Q\/b)-1\right\}\,,
\label{inbq}
\end{equation}
and the function $\bar f_{NS}(n,b,Q_0)$ is given by the input unintegrated distribution
at the reference scale $Q_0$.
At $b=0$ we have:
\bea
\tilde f_{NS}(n,b=0,Q)\!\!&=&\!\!{1\over 2}\,p_{NS}(n,Q^2)
\\ \nonumber
\\
\tilde f_{NS}(n,b=0,Q_0)\!\!&=&\!\!{1\over 2}\,p_{NS}(n,Q_0^2)
\label{unvin}
\eea
where $p(n,Q^2)$ are the moment functions of the integrated distributions.
We shall now rearrange  the integral $I(n,b,Q)$ as below:
\begin{equation}
I(n,b,Q)\,=\,I_1(n,b,Q)\,+\,I_2(b,Q)
\label{i12}
\end{equation}
where
\bea
I_1(n,b,Q)
\!\!&=&\!\!
\int_{Q_0^2}^{Q^2}{dq^{2}\over q^{2}}
{\alpha_s(q^{2})\over 2 \pi}\int_0^1 dz\, P_{qq}(z)\,(z^n-1)\,J_0((1-z)\/Q\/b)
\label{inbq1}
\\ \nonumber
\\
I_2(b,Q)
\!\!&=&\!\!
\int_{Q_0^2}^{Q^2}{dq^{2}\over q^{2}}
{\alpha_s(q^{2})\over 2 \pi}\int_0^1 dz\, P_{qq}(z)\left\{J_0((1-z)\/Q\/b)-1\right\}\,.
\label{ibq2}
\eea

The dominant contribution to the integrals $I_1$ and $I_2$ comes from the
region in which argument of the Bessel function $J_0(u)$ is small,  where
the Bessel function itself can be approximated by unity, i.e. we can
adopt the following  approximation:
\begin{equation}
J_0(u)\, \simeq\, \Theta(c-u)
\label{j01}
\end{equation}
with $c$ being of the order of 1; for simplicity we  set $c=1$.
The dominant contribution to the integral $I_1$ comes from the region $Q\/b<1$
and adopting approximation (\ref{j01}) we get for $b>1/Q$:
\begin{equation}
I_1(n,b,Q)\, \simeq \,\int_{Q_0^2}^{1/b^2}{dq^{ 2}\over q^{ 2}}
{\alpha_s(q^{2})\over 2 \pi}\int_0^1 dz\, P_{qq}(z)\,(z^n-1)\,.
\label{inb1a}
\end{equation}
 We observe that approximate representation (\ref{inb1a}) of $I_1(n,b,Q)$
is equal to the argument of the exponent
describing evolution of the moments of the {\it integrated} non-singlet quark
distributions
from the scale $Q_0^2$ to the scale $1/b^2$.  For $b\ll1/Q_0$ we may set
\beq
\tilde f_{NS}(n,b,Q_0)\, \simeq\,\tilde f_{NS}(n,b=0,Q_0)\, =\, {1\over 2}\,\tilde p_{NS}(n,Q_0^2)
\eeq
to finally get
\begin{equation}
\tilde f_{NS}(n,b,Q_0)\,\exp\{I_1(n,b,Q)\}\, \simeq\, {1\over 2}\,\tilde  p_{NS}\!\left(n,{1\over b^2}\right)\,.
\label{fact1}
\end{equation}
After inverting moments one finds
\beq
\bar f_{NS}(x,b,Q)\,\simeq\, {1\over 2}\,x\,p_{NS}(x,1/b^2)\,.
\eeq
 Comparing
(\ref{wij}) with (\ref{swij}) we find
that in the functions $W_{ij}(x_1,x_2,b,Q)$ defined by the non-singlet quark distributions,
we have  identified
the first factor in the soft gluon resummation formula to be equal to the product
of the integrated
parton distributions at the scale of the order of $1/b^2$.

We now show that after using approximation
(\ref{j01}) in the evaluation of $I_2(b,Q)$,  the form factor
$\exp\{2I_2(b,Q)\}$  becomes equal to the  form factor
$\exp\{S_q(b,Q)\}$ in the LO approximation
with the exponent $S_q(b,Q)$ defined by eq.~(\ref{sibq}).
To this aim we note that after rearranging the splitting function $P_{qq}(z)$
\begin{equation}
P_{qq}(z)\,=\,C_F\left[{2\over 1-z}-(1+z)\right]\,,
\label{pqqspl}
\end{equation}
we can represent the integral $I_2(b,Q)$ as the sum of two terms:
\begin{equation}
I_2(b,Q)\,=\,I_2^1(b,Q)\,+\,I_2^2(b,Q)
\label{isplit2}
\end{equation}
where
\bea
I_2^1(b,Q)
\!\!&=&\!\!
2\,C_F\int_{Q_0^2}^{Q^2}{dq^{2}\over q^{2}}
{\alpha_s(q^{2})\over 2 \pi}\int_0^1 {dz \over 1-z}\left[J_0((1-z)q\/b)-1\right]
\label{i21}
\\ \nonumber
\\
I_2^2(b,Q)
\!\!&=&\!\!
-C_F\int_{Q_0^2}^{Q^2}{dq^{2}\over q^{2}}
{\alpha_s(q^{2})\over 2 \pi}\int_0^1 dz\,  (1+z)\left[J_0((1-z)q\/b)-1\right]\,.
\label{i22}
\eea
Using approximation (\ref{j01}) we get:
\bea
I_2^1(b,Q)
\!\!&\simeq&\!\!
-\Theta(Q^2-1/b^2)\,C_F\int_{1/b^2}^{Q^2}
{dq^2\over q^2}{\alpha_s(q^2)\over 2 \pi}\, \ln(q^2\/b^2)
\label{i21a}
\\ \nonumber
\\ \nonumber
I_2^2(b,Q)
\!\!&\simeq&\!\!
\Theta(Q^2-1/b^2)\,C_F\int_{1/b^2}^{Q^2}
{dq^2\over q^2}{\alpha_s(q^2)\over 2 \pi}\bigg(1-{1\over bq}\bigg)
\bigg(1+{1\over 2}\bigg(1-{1\over bq}\bigg)\bigg)
\\ \nonumber
\\
\!\!&\simeq&\!\!
\Theta(Q^2-1/b^2)\,{3\over 2}\,C_F\int_{1/b^2}^{Q^2}
{dq^2\over q^2}{\alpha_s(q^2)\over 2 \pi}\,.
\label{i22a}
\eea
Thus (\ref{isplit2}) becomes
\begin{equation}
2\,I_2(b,q)\, \simeq\, -2\,\Theta(Q^2-1/b^2)\,C_F\,\int_{1/b^2}^{Q^2}
{dq^2\over q^2}{\alpha_s(q^2)\over 2 \pi} \left[\ln(q^2\/b^2)-{3\over 2}\right]\,.
\label{i2fin}
\end{equation}
We note that this expression coincides with that  given by eq.~(\ref{sibq})
defing the exponent $S_q(b,Q)$ in the LO approximation. To be precise, the first term
corresponding to $I_2^1(b,Q)$ is identically equal to the term defined by $A_q(\alpha_s)$ for the
fixed coupling only and there are subleading differences due to different scale of the QCD
coupling. It may be observed, however, that we would get exactly the same result for this term choosing
the scale of the coupling equal to $(q(1-z))^2$ instead of $q^2$ in the integrals
defining $I_2^1(b,Q)$.  The second term corresponding to $I_2^2(b,Q)$ is identical to
that defined by $B_q(\alpha_s)$.
Finally we observe that the non-perturbative
contribution factorises as the $Q$ independent factor, see eq. (\ref{smomns}).


The valence quarks dominate in the production of DY muon pairs and in
the production of gauge $W^{\mp}$ or $Z^{0}$ bosons in $p \bar p$ collisions.
However, inclusive production of the Higgs bosons, which is dominated by the gluon-gluon fusion,
requires the knowledge of gluon and sea quark distributions.
We shall now derive the soft gluon resummation
formula for the unintegrated gluon and sea quark distributions, and for the function
$W_{GG}(x_1,x_2,b,Q)$, starting from the system of the CCFM equations (\ref{dccfmsb},\ref{dccfmgb}).
In this case we cannot solve the system of the CCFM equations in an analytic form and the
soft gluon resummation expressions
will follow from the approximate treatment of the CCFM equations themselves.
To this aim it is convenient to integrate
these equations and represent them as the
system of integral equations:
\bea
\label{ccfmgb0}
\nonumber
\bar f_{S}(x,b,Q)
\!\!&=&\!\!
\bar f_S^0(x,b)
+ \int_0^1dz\int{dq^2\over q^2} {\alpha_s(q^2)\over 2\pi}\,\Theta(q^2-q_0^2)\,\Theta(Q-q)
\\ \nonumber
\\ \nonumber
\!\!&\times&\!\!
\bigg\{\Theta(z-x)\,J_0((1-z)qb)\left[P_{qq}(z)\,\bar f_{S}\!\left({x\over z},b,q\right)+
P_{qg}(z)\,\bar f_{G}\!\left({x\over z},b,q\right)\right]
\\ \nonumber
\\
&&\,\,\,\,- \,
[zP_{qq}(z)+zP_{gq}(z)]\,\bar f_{S}(x,b,q)\bigg\}
\eea
\bea
\nonumber
\bar f_{G}(x,b,Q)
\!\!&=&\!\!
\bar f_G^0(x,b)
+ \int_0^1dz\int{dq^2\over q^2} {\alpha_s(q^2)\over 2\pi}\,\Theta(q^2-q_0^2)\,\Theta(Q-q)
\\ \nonumber
\\ \nonumber
\!\!&\times&\!\!
\bigg\{\Theta(z-x)\,J_0((1-z)qb)\left[P_{gq}(z)\,\bar f_{S}\!\left({x\over z},b,q\right)+
P_{gg}(z)\,\bar f_{G}\!\left({x\over z},b,q\right)\right]
\\ \nonumber
\\
&&\,\,\,\,- \,
[zP_{gg}(z)+zP_{qg}(z)]\,\bar f_{G}(x,b,q)\bigg\}\,.
\label{ccfmgb}
\eea
We rearrange these equations by adding and subtracting in the integrand of (\ref{ccfmgb0}) the term
\beq
\nonumber
{\alpha_s(q^2)\over 2\pi q^2}\,[zP_{qq}(z)+zP_{gq}(z)]\,J_0\left((1-z)bq\right)\bar f_{S}(x,b,q)\,,
\eeq
and similarly in  the integrand of (\ref{ccfmgb}) the term
\beq
\nonumber
{\alpha_s(q^2)\over 2\pi q^2}\,[zP_{gg}(z)+zP_{qg}(z)]\,J_0\left((1-z)bq\right)\bar f_{G}(x,b,q)\,.
\eeq
Thus we obtain
\bea
\nonumber
\bar f_{S}(x,b,Q) \!\!&=&\!\! \bar f_S^0(x,b)
\,+\, \int_0^1dz\int{dq^2\over q^2} {\alpha_s(q^2)\over 2\pi}\,\Theta(q^2-q_0^2)\,\Theta(Q-q)
\\ \nonumber
\\ \nonumber
&\times&\!\!
\bigg\{
J_0((1-z)qb)\bigg[\Theta(z-x)
\left[
P_{qq}(z)\,\bar f_{S}\!\left({x\over z},b,q\right)
+P_{qg}(z)\,\bar f_{G}\!\left({x\over z},b,q\right)
\right]
\\ \nonumber
\\ \nonumber
&&\,\,\,\,- \left[zP_{qq}(z)+zP_{gq}(z)\right]\,\bar f_{S}(x,b,q)\bigg]
\\ \nonumber
\\
&&\,\,\,\,- \left[zP_{qq}(z)+zP_{gq}(z)\right]\,\left[1-J_0((1-z)qb)\right]\,\bar f_{S}(x,b,q)\bigg\}
\label{ccfmsb1}
\eea
\bea
\nonumber
\bar f_{G}(x,b,Q) \!\!&=&\!\!  \bar f_G^0(x,b)
\,+\, \int_0^1dz\int{dq^2\over q^2} {\alpha_s(q^2)\over 2\pi}\,\Theta(q^2-q_0^2)\,\Theta(Q-q)
\\ \nonumber
\\ \nonumber
&\times&\!\! \bigg\{J_0((1-z)qb)\bigg[\Theta(z-x)
\left[P_{gq}(z)\,\bar f_{S}\left({x\over z},b,q\right)
+P_{gg}(z)\,\bar f_{G}\left({x\over z},b,q\right)
\right]
\\ \nonumber
\\ \nonumber
&&\,\,\,\, - \left[zP_{gg}(z)+zP_{qg}(z)\right]\, \bar f_{G}(x,b,q)\bigg]
\\ \nonumber
\\
&&\,\,\,\, - \left[zP_{gg}(z)+zP_{qg}(z)\right] [1-J_0((1-z)qb)]\,\bar f_{G}(x,b,q)\bigg\}
\label{ccfmgb1}
\eea

We consider this equation in the region $b\ll 1/Q_0$, set $b=0$
in the inhomogeneous term,  make the approximation
$\{1-J_0((1-z)qb))\} \simeq \Theta((1-z)qb-1)$ in the last terms and limit the integration
to $q<min\{Q,1/b\}$ in the remaining terms, to get for $b>1/Q$ :
\bea
\nonumber
\bar f_{S}(x,b,Q)
\!\!&=&\!\!
{1\over 2}\, x\/q_S(x,Q_0^2)
\,+\, \int{dq^2\over q^2} \int_0^1dz\,{\alpha_s(q^2)\over 2\pi}\,\Theta(q^2-q_0^2)\bigg\{\Theta(1/b-q)
\\ \nonumber
\\ \nonumber
\!\!&\times&\!\!
\bigg[\Theta(z-x)\!\left[
P_{qq}(z)\,\bar f_{S}\left({x\over z},b,q\right)
+P_{qg}(z)\,\bar f_{G}\left({x\over z},b,q\right)
\right]
\\ \nonumber
\\ \nonumber
&&~
-[zP_{qq}(z)+zP_{gq}(z)]\,\bar f_{S}(x,b,q)\bigg]
\\ \nonumber
\\
\!\!&-&\!\!
\Theta(Q-q)\,\Theta(q-1/b)\,\Theta(1-1/(qb)-z)\,
[zP_{qq}(z)+zP_{gq}(z)]\bar f_{S}(x,b,q)\bigg\}
\label{ccfmsb2}
\eea
\bea
\nonumber
\bar f_{G}(x,b,Q)
\!\!&=&\!\!
{1\over 2}\, x\/g(x,Q_0^2)
\,+\, \int{dq^2\over q^2} \int_0^1dz\,{\alpha_s(q^2)\over 2\pi}\,\Theta(q^2-q_0^2)\bigg\{\Theta(1/b-q)
\\ \nonumber
\\ \nonumber
\!\!&\times&\!\!
\bigg[\Theta(z-x)\!\left[
P_{gq}(z)\,\bar f_{S}\!\left({x\over z},b,q\right)
+P_{gg}(z)\,\bar f_{G}\!\left({x\over z},b,q\right)
\right]
\\ \nonumber
\\ \nonumber
&&~
-[zP_{gg}(z)+zP_{qg}(z)]\,\bar f_{G}(x,b,q)\bigg]
\\ \nonumber
\\
\!\!&-&\!\!
\Theta(Q-q)\,\Theta(q-1/b)\,\Theta(1-1/(qb)-z)\,
[zP_{gg}(z)+zP_{qg}(z)]\,\bar f_{G}(x,b,q)\bigg\}
\label{ccfmgb2}
\eea
where
\beq
q_S(x,Q_0^2)\,=\,
\sum_{i=1}^{N_f}(q_i(x,Q_0^2)+\bar q_i(x,Q_0^2))\,.
\eeq
We note that $f_i(x,b,q)$ and $f_i(x/z,b,q)$ in all terms, except those in the last terms of
eqs.~(\ref{ccfmsb2}) and (\ref{ccfmgb2}),  are in the region $q<1/b$.  We shall show that those terms give
$1/2\,xq_{S}(x,1/b^2)$ and $1/2\,xg(x,1/b^2)$, respectively.
Let us observe at first that  for $Q<1/b$ eqs.~(\ref{ccfmsb2}) and  (\ref{ccfmgb2})
read:
\bea
\nonumber
\bar f_{S}(x,b,Q)
\!\!&=&\!\!
{1\over 2}\, x\/q_S(x,Q_0^2)
\,+\, \int_0^1dz\int{dq^2\over q^2} {\alpha_s(q^2)\over 2\pi}\,\Theta(q^2-q_0^2)\,\Theta(Q-q)
\\ \nonumber
\\ \nonumber
&\times&\!\!
\bigg\{\Theta(z-x)\left[
P_{qq}(z)\,\bar f_{S}\!\left({x\over z},b,q\right)
+P_{qg}(z)\,\bar f_{G}\!\left({x\over z},b,q\right)
\right]
\\ \nonumber
\\
&&~-
[zP_{qq}(z)+zP_{gq}(z)]\,\bar f_{S}(x,b,q)\bigg\}
\label{ccfmsb3}
\eea
\bea\nonumber
\bar f_{G}(x,b,Q)
\!\!&=&\!\!
{1\over 2}\, x\/g(x,Q_0^2)
\,+\, \int_0^1dz\int{dq^2\over q^2} {\alpha_s(q^2)\over 2\pi}\,\Theta(q^2-q_0^2)\,\Theta(Q-q)
\\ \nonumber
\\ \nonumber
&\times&\!\!
\bigg\{\Theta(z-x)\left[
P_{gq}(z)\,\bar f_{S}\!\left({x\over z},b,q\right)
+P_{gg}(z)\,\bar f_{G}\!\left({x\over z},b,q\right)
\right]
\\ \nonumber
\\
&&~-
[zP_{gg}(z)+zP_{qg}(z)]\,\bar f_{G}(x,b,q)\bigg\}\,,
\label{ccfmgb3}
\eea
which are just the DGLAP equations in the integral form for $1/2\,xg(x,Q^2)$ and $1/2\,xq_S(x,Q^2)$.
We therefore get
$f_G(x,b)=1/2\,xg(x,Q^2)$ for $b<1/Q$.  The first terms in eq.~(\ref{ccfmsb2}) and
(\ref{ccfmgb2}) are the same as in the
right hand side of equations (\ref{ccfmsb3}) and (\ref{ccfmgb3})  except that instead of $\Theta(Q-q)$
we have $\Theta(1/b-q)$.  The first terms in eqs.~(\ref{ccfmsb2}) and (\ref{ccfmgb2})
reduce to $1/2\,xq_S(x,1/b^2)$ and
$1/2\,xg(x,1/b^2)$. Therefore,  eqs.~(\ref{ccfmsb2}) and (\ref{ccfmgb2}) read:
\bea
\nonumber
\bar f_{S}(x,b,Q)
\!\!&=&\!\!
{1\over 2}\, x\/q_S(x,1/b^2)
- \int_0^1dz\int{dq^2\over q^2} {\alpha_s(q^2)\over 2\pi}
\\ \nonumber
\\
\!\!&\times&\!\!
\Theta(q-1/b)\,\Theta(Q-q)\,\Theta(1-1/(qb)-z)\,
[zP_{qq}(z)+zP_{gq}(z)]\,\bar f_{S}(x,b,q)
\label{ccfmgb4a}
\eea
\bea\nonumber
\bar f_{G}(x,b,Q)
\!\!&=&\!\!
{1\over 2}\, x\/g(x,1/b^2)
- \int_0^1dz\int{dq^2\over q^2} {\alpha_s(q^2)\over 2\pi}
\\ \nonumber
\\
\!\!&\times&\!\!
\Theta(q-1/b)\,\Theta(Q-q)\,\Theta(1-1/(qb)-z)\,
[zP_{gg}(z)+zP_{qg}(z)]\,\bar f_{G}(x,b,q)\,,
\label{ccfmgb4b}
\eea
and their solution read:
\bea
\bar f_S(x,b,Q)
\!\!&=&\!\!
{1\over 2}\,\exp\{1/2\,\tilde S_q(b,Q)\}\,xq_S(x,1/b^2)
\\ \nonumber
\\
\bar f_G(x,b,Q)
\!\!&=&\!\!
{1\over 2}\,\exp\{1/2\,\tilde S_G(b,Q)\}\,xg(x,1/b^2)
\label{cssg}
\eea
with $\tilde S_q$ and $\tilde S_G$ defined by:
 \bea
\tilde S_q(b,Q)
\!\!&=&\!\!
-2\int_{1/b^2}^{Q^2}{dq^2\over q^2} {\alpha_s(q^2)\over 2\pi}
\int_0^{1-1/(bq)}dz\,
[zP_{qq}(z)+zP_{gq}(z)]
\label{tilsq}
\\ \nonumber
\\
\tilde S_G(b,Q)
\!\!&=&\!\!
-2\int_{1/b^2}^{Q^2}{dq^2\over q^2} {\alpha_s(q^2)\over 2\pi}
\int_0^{1-1/(bq)}dz\,
[zP_{gg}(z)+zP_{qg}(z)]
\label{tilsg}
\eea
Substituting explicit expressions for the splitting functions $P_{ij}(z)$ and
setting $1$ instead of $1-1/(bq)$ as the upper integration limit in the integrals over
non-singular at $z=1$ terms,  and using eqs.~(\ref{rg1})-(\ref{rqpqq}) we get:
\begin{equation}
\tilde S_G(b,Q)\, \simeq\, -\int_{1/b^2}^{Q^2}{dq^2\over q^2} {\alpha_s(q^2)\over 2\pi}\,
\left[A_G^{(1)}\ln(q^2b^2)+B_G^{(1)}\right]
\label{tilsg1}
\end{equation}
with $A_G^{(1)}$ and $B_G^{(1)}$ given by eq.~(\ref{log}).
We also find that $\tilde S_q(b,Q)$ is given by eq.~(\ref{i2fin}).
As in the case of non-singlet quarks the  first term on the right hand side
of eq.~(\ref{tilsg1}), with  the
integrand  proportional to $\ln(q^2b^2)$, gives the result which is given by
the  term with $A_G(\alpha_s)$ in eq.~(\ref{sibq}) and the
fixed coupling only.  Thus,  the subleading differences are due to a different scale of  the QCD
coupling.  We would get exactly the same result for this term choosing
the scale of $\alpha_s$ equal to $[q(1-z)]^2$ instead of $q^2$ in the integrand
defining $I_2^1(b,Q)$, eq.~(\ref{i21}).  
The second term on the right hand side of eq.~(\ref{tilsg1}) is identical to
that defined by $B_G(\alpha_s)$ in eq.~(\ref{sibq}).
As in the case of the valence quark distributions also for the gluon and sea quark
distributions the non-perturbative effects will enter as the $Q-$independent factor(s).

\section{Numerical results}

We have shown in the previous sections that the parton shower described by the
CCFM equations in the single loop approximation (\ref{dccfmnsb})-(\ref{dccfmgb})
generate complete  resummation effects in LO, i.e. it
automatically resums the double and single ``large'' logarithms $\ln^2(Q^2b^2)$ in the
region $Q^2b^2\gg 1$.  In this Section we  present numerical results based on the
exact (numerical)  solution of the CCFM equations in the $b-$representation and compare
them with the predictions based on the resummation formulas.
In our analysis we assume   a factorisable form
of the initial conditions (\ref{bcond})
\begin{equation}
\bar f_k^0(x,b)\,=\,{1\over 2}\,F(b)\,x\/p_k(x,Q_0^2)\,,
\label{bcondf}
\end{equation}
assuming for simplicity the same input profile $F(b)$ for quarks and gluons. It is determined
by a
non-perturbative $k_T-$distribution at the scale $Q_0$, and  at $b=0$ we have
the normalisation condition $F(0)=1$.  Thus, we assume the Gaussian
form of the profile function
\begin{equation}
F(b)\,=\, \exp\left\{-{b^2/b_0^2}\right\}\,,
\label{fbgauss}
\end{equation}
where we  set $b_0^2=4\,\mbox{\rm GeV}^{-2}$.  The starting integrated distributions $p_k(x,Q_0^2)$
are taken from the LO GRV analysis \cite{GRV}. With the factorisable form (\ref{bcondf})
of the starting quark and gluon distributions, we obtain
the corresponding factorisation of the function (\ref{wij})
\begin{equation}
W_{ij}(x_1,x_2,b,Q)\,=\,F^2(b)\,\tilde W_{ij}(x_1,x_2,b,Q)\,,~~~~~~i,j=q,{\bar{q}},g\,.
\label{wwt}
\end{equation}
For the soft gluon resummation,
the  expression for the function  ${\tilde W_{ij}}$ takes the following form
\begin{equation}
\tilde W^{resum}_{ij}(x_1,x_2,b,Q)\,=\,\exp\{1/2\,[S_i(b,Q)+S_j(b,Q)]\}\,x_1p_i(x_1,1/b^2)\,
\,x_2p_j(x_2,1/b^2)\,.
\label{twresum}
\end{equation}

 In Fig. 1 we show
results for $\tilde W_{gg}(x_1,x_2,b,Q=M_H)$ plotted as the
function of $b^2$,   calculated for $x_1=x_2=M_H/\sqrt{s}$ for
$\sqrt{s}$ equal to  1.8 TeV (Fig.~1a)  and 14 TeV (Fig.~1b).  
The Higgs boson mass was assumed to be equal to 115
GeV.  We show the results based upon the numerical solution of the CCFM equations
(\ref{dccfmnsb})-(\ref{dccfmgb}) (solid lines)
and confront them with the predictions based upon the soft gluon
resummation formula with the argument of the exponent $S_G(b,Q)$
given by eq.~(\ref{tilsg1}) (dotted lines). We also show the results
corresponding to $S_G(b,Q)$ given by the standard eq.~(\ref{sibq}) (dashed lines). In
both latter cases the scale $1/b^2$ was replaced by
$Q^2/(b^2Q^2+1)$ that guarantees smooth transition to the region
$b^2<1/Q^2$.  We have also replaced $S_G(b,Q)$ by $S_G^{eff}(b,Q)$,
setting $S_G^{eff}(b,Q)=0$ in the region where $S_G(b,Q)>0$, and
$S_G^{eff}(b,Q)=S_G(b,Q)$ in the region where $S_G(b,Q)<0$.  We
can see that the function $W_{gg}$ calculated from the exact
solution of the CCFM equation stays most of the time below the
predictions based upon the resummation formulas.  Only at
relatively large values of $b \sim 1~\mbox{\rm GeV}^{-1}$, the
resummation formula corresponding to $S_G(b,Q)$ given by eq.~(\ref{tilsg1}) 
leads to smaller magnitude of $W_{gg}$ than that
corresponding to the exact solution of the CCFM equation.  This
difference is caused by a different behaviour of the
Sudakov-like form factors $T_g=\exp\{S_G(b,Q)\}$ in all three cases, as
can be seen in Fig.~2. In this figure we plot as the solid line the function
$\exp\{S_G^{CCFM}(b,Q)\}$ with
\begin{equation}
S_G^{CCFM}(b,Q)\,=\,\int_{q_0^2}^{Q^2}{dq^2\over q^2} {\alpha_s(q^2)\over 2 \pi}\int_0^1 dz\,
z\,[P_{gg}(z)+P_{qg}(z)]\,\{J_0((1-z)qb)-1\}\,,
\label{sccfm}
\end{equation}
and compare it with the functions resulting from the soft resummation
functions with $S_G$ given by
eq.~(\ref{tilsg1})  (dotted line) and eq.~(\ref{sibq}) (dashed line).

In Fig.~3
we show the impact of those differences on
 the transverse momentum  distribution of Higgs bosons produced at $y=0$, eq.~(\ref{sighb1}).
We present the results for  $\sqrt{s}=1.8~ TeV$ and for $\sqrt{s}=14~TeV$.   
 We  can see that the results based upon the exact
solution of the CCFM equation (solid lines) gives smaller magnitude of the
cross section at its maximum that those generated by the soft gluon
resummation formulas (\ref{tilsg1}) (dotted lines) and (\ref{sibq}) (dashed lines).
The former tends, however, to
generate the longer tail of the $p_T$ distribution.

\section{Summary and conclusions}

In this paper we have discussed the transverse momentum distributions of the Higgs bosons
within the unintegrated distributions framework.  We considered the $gg\rightarrow H$ fusion
process which is the dominant mechanism of inclusive Higgs boson production.
The unintegrated distributions were obtained from the CCFM equations in the single loop
approximation which is equivalent to the LO DGLAP evolution for the integrated distributions.  We have
utilised the fact that the CCFM equations in the single loop approximation can be
diagonalised in the $b$ representation.  We have shown  that the conventional
soft gluon resummation formulas are embodied in the solution(s) of the CCFM equations.
We have examined possible differences between the exact (numerical) solution of the CCFM equation
and their approximate solution leading to conventional soft gluon resummation formulas.

\section*{Acknowledgments}
This research was partially supported
by the Polish Committee for Scientific Research (KBN) grant no. 5P03B 14420.
This paper was completed before Jan Kwieci\'nski passed away. The help of
Krzysztof Golec--Biernat in preparation of the manuscript is kindly acknowledged.


\newpage
\begin{figure}
\begin{center}
\includegraphics*[angle=-90,width=0.7\textwidth]{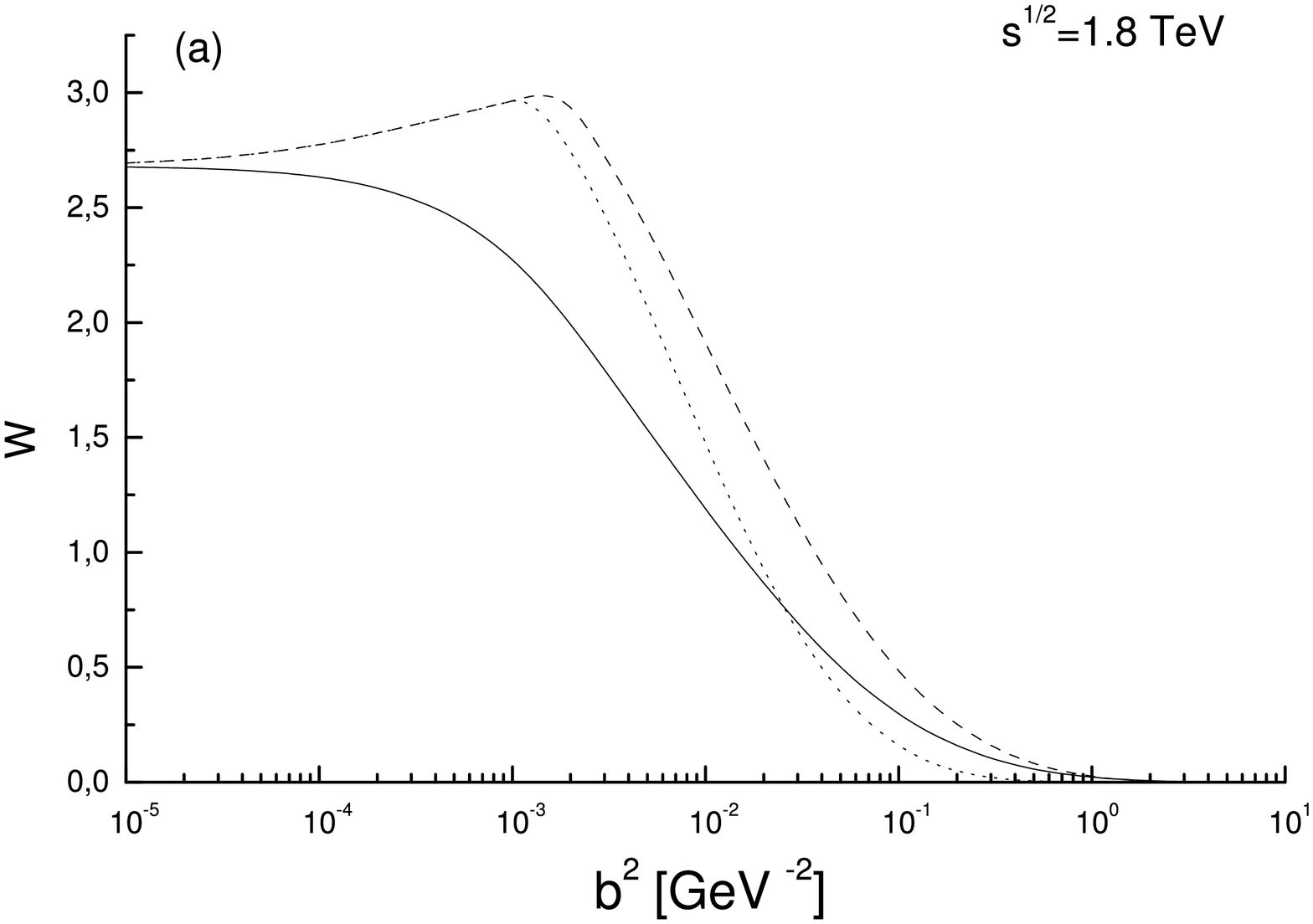}
\end{center}
\end{figure}
\begin{figure}
\begin{center}
\includegraphics*[angle=-90,width=0.7\textwidth]{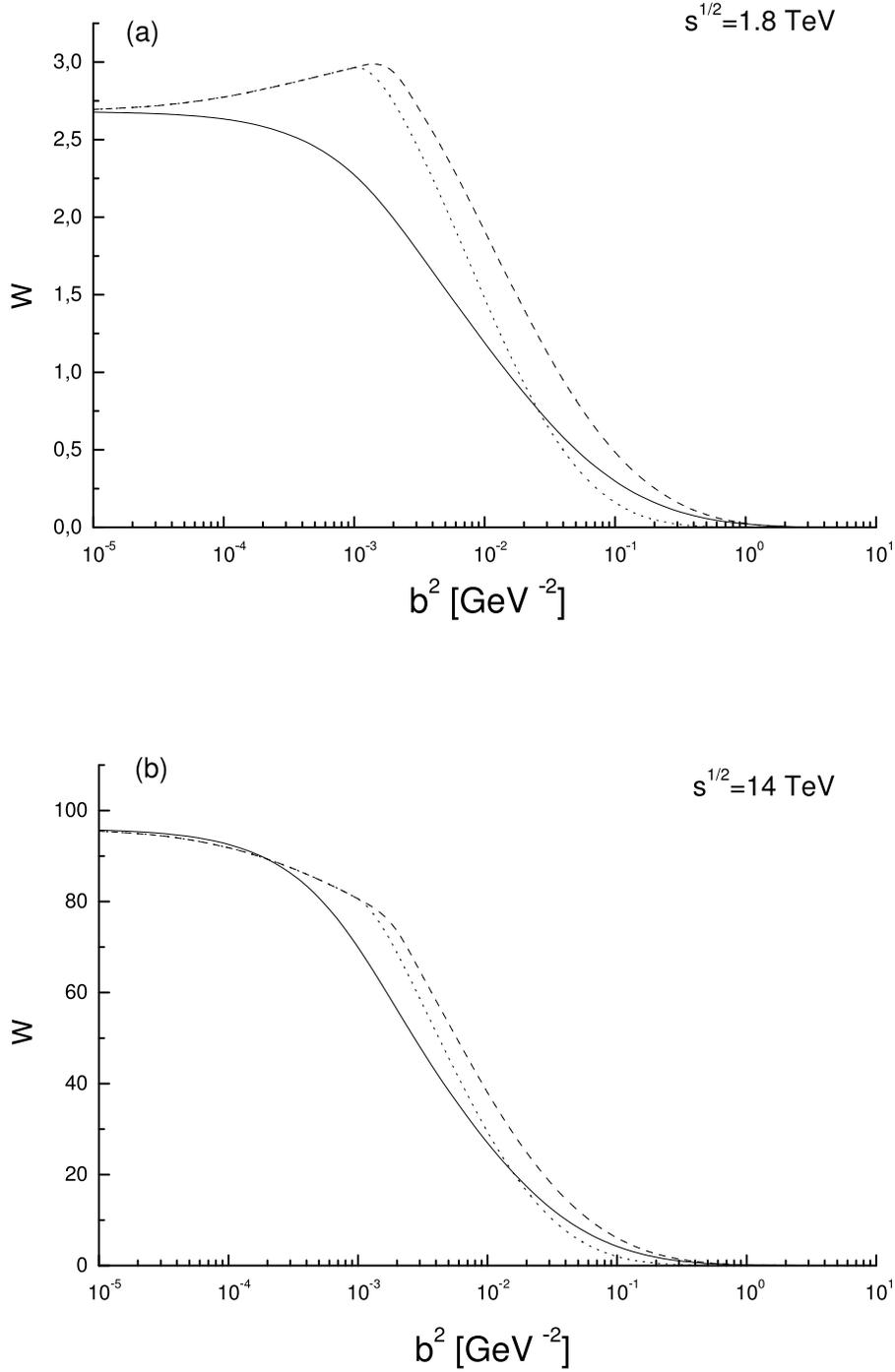}
\caption{The function $W_{GG}(x_1,x_2,b,Q)$ as a function of $b$ for the Higgs boson production
at $Q=m_H=115~\mbox{\rm  GeV}$ in the gluon - gluon fusion at the Tevatron (a) and the  LHC energy (b).
The solid lines correspond to the result with the unintegrated gluon distributions
from the CCFM equations.  The results with the LO soft gluon resumation formulas are
shown by the dotted lines ($S_G$ given by eq.~(\ref{tilsg1})), and the 
dashed lines ($S_G$ given by eq.~(\ref{sibq})).
}
\end{center}
\end{figure}

\newpage

\begin{figure}
\begin{center}
\includegraphics*[angle=-90,width=0.7\textwidth]{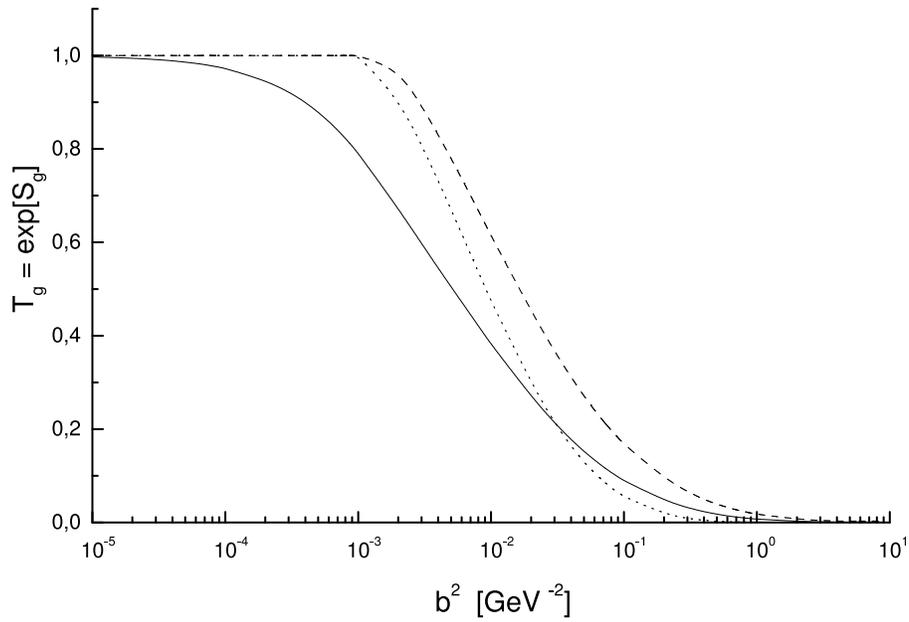}
\vskip 1cm
\caption{The Sudakow-like form  factor  $T_G=\exp\{S_G\}$,
as a function of the impact parameter $b$, from the CCFM equations (solid line)
and the soft gluon resummation formulas (dotted and dashed lines). The function $S_G$ is
given by eq.~(\ref{sccfm}) for the solid line, eq.~(\ref{tilsg1})
for the dotted line, and eq.~(\ref{sibq}) for the dashed line.
}
\end{center}
\end{figure}

\newpage

\begin{figure}
\begin{center}
\includegraphics*[angle=-90,width=0.7\textwidth]{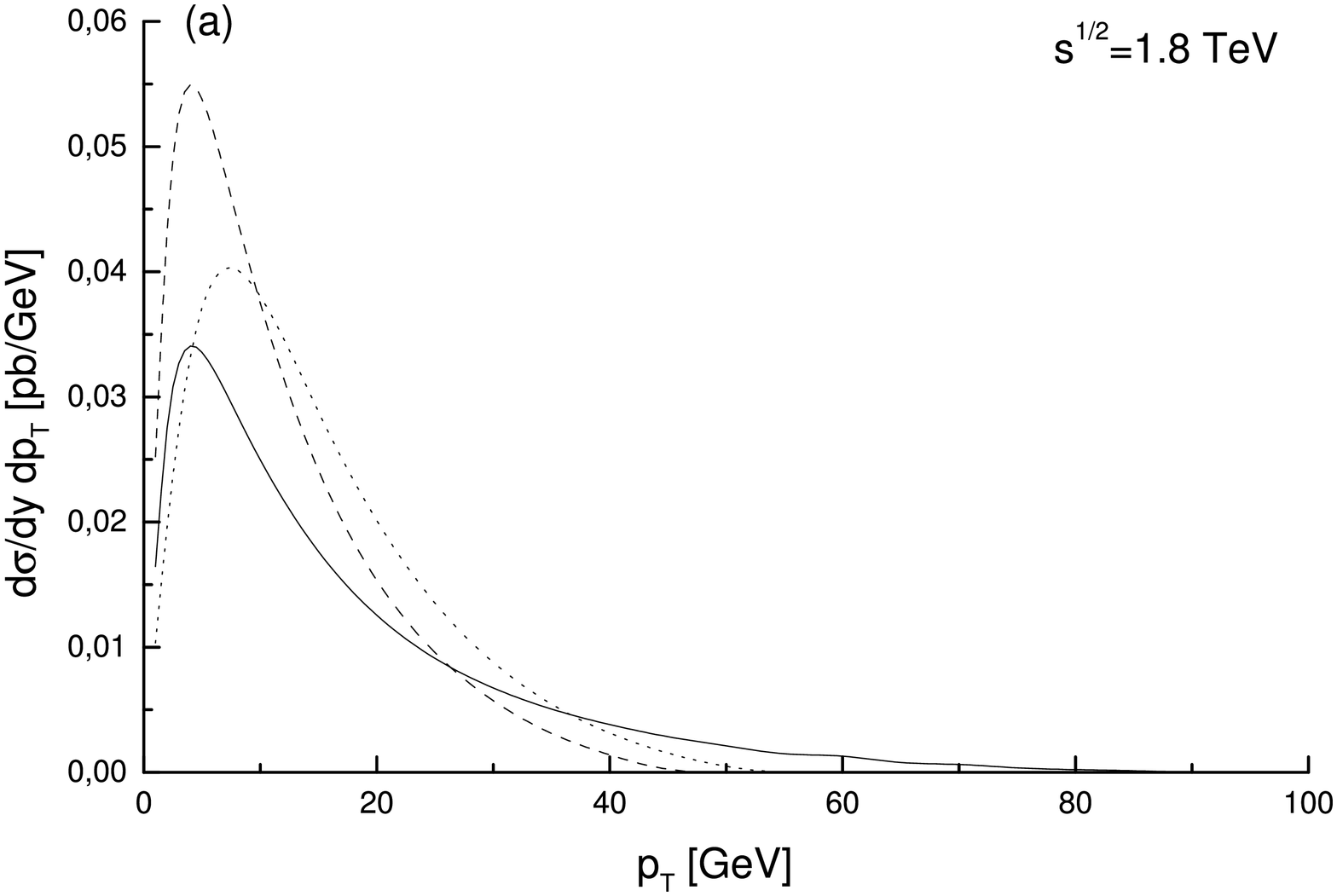}
\vskip 2cm
\includegraphics*[angle=-90,width=0.7\textwidth]{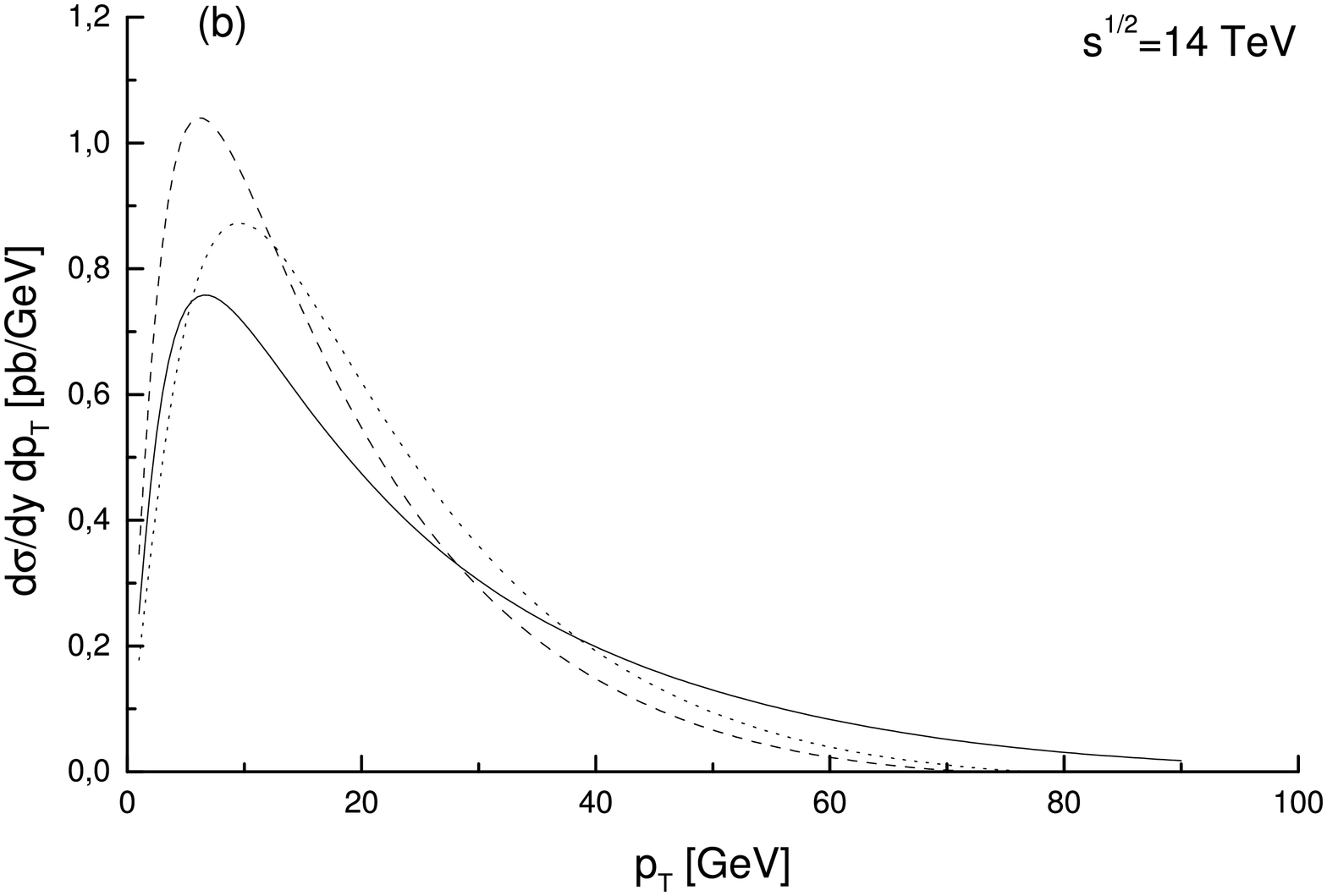}
\caption{
The transverse momentum distribution  of Higgs boson from  the gluon-gluon fusion
at $Q=m_H=115~\mbox{\rm GeV}$ for central rapidity, $y=0$.
Figures (a) and (b) correspond to the Tevatron and LHC
energy, respectively. 
The solid lines are based on the numerical solution of the CCFM equations
(\ref{dccfmnsb})-(\ref{dccfmgb}), while the the dotted and dashed lines 
results from the LO soft gluon resummation formulas with $S_G$
given by eq.~(\ref{tilsg1}) and eq.~(\ref{sibq}), respectively.
}
\end{center}
\end{figure}
\end{document}